\documentclass{ws-procs9x6}

\begin{document}

\title{Positivity of QCD at asymptotic density
}

\author{Deog Ki Hong}

\address{Department of Physics, Pusan National University,\\
Busan 609-735, Korea\\
E-mail: dkhong@pusan.ac.kr}


\maketitle

\abstracts{
In this talk, I try to show that the sign problem of dense
QCD is due to modes whose frequency is higher than the chemical potential.
An effective theory
of quasi-quarks near the Fermi surface has a positive measure
in the leading order. The higher-order corrections
make the measure complex, but
they are suppressed as long as the chemical potential is
sufficiently larger than $\Lambda_{\rm QCD}$. As a consequence
of the positivity of the effective theory, we can show that
the global vector symmetries except the $U(1)$ baryon number are
unbroken at asymptotic density.
}

\section{Introduction}
It is now firmly believed that quantum chromodynamics (QCD)
successfully describes the strong interaction. Its prediction
on the hadron interaction at high energy is well confirmed
by experiments. The coupling
extracted from various hadronic processes scales
logarithmically with
respect to energy as QCD predicted~\cite{Gross:1973id,Politzer:1973fx}.
Furthermore, the low energy dynamics of hadrons is
in good agreement of the chiral symmetry breaking of
QCD~\cite{Gasser:1983yg}.
It explains successfully why pions and
kaons are much lighter than baryons and why hadron spectra are
not in parity-doublet, though the strong interaction preserves parity.

Being the theory of strong interaction, QCD should be able to
tell us how matter
behaves at extreme environments, as encountered in heavy ion
collisions, in early universe, or in compact stars. One salient
feature of dense matter is that it undergoes
phase transitions at extreme environments. QCD predicts
phase transitions for hot and dense matter as one increases the
temperature or the density of matter. On dimensional ground, the
critical temperature and the critical chemical potential have to be
of order of $\Lambda_{\rm QCD}$, which is the only dimensional parameter
that QCD dynamics depends on; $T_C\sim \Lambda_{\rm QCD}$
and $\mu_C\sim\Lambda_{\rm QCD}$. In fact, some of these predictions
on hot matter have been confirmed by lattice calculations~\cite{Karsch:2000kv}.
Lattice calculation shows $T_C=175~{\rm MeV}$, which is close
to $\Lambda_{\rm QCD}(\simeq 213~{\rm MeV})$. However, little progress
in lattice QCD has been made  to probe the density phase transition in
matter except when the chemical
potential is small~\cite{Fodor:2001au},
since lattice QCD at finite density suffers a notorious
sign problem due the complexity of the measure~\cite{Hands:2001jn}.

In this talk, I will argue that the sign problem can be solved
for certain quantities at high density,
thus allowing lattice calculation, and the QCD measure
becomes positive at asymptotic density.

\section{High density effective theory}

Quark matter on lattice is described by a partition function given as
\begin{equation}
Z(\mu)=\int {\rm d}A~\det \left(M\right)e^{-S(A)},
\end{equation}
where $M=\gamma_E^{\mu}D_E^{\mu}+\mu\gamma_E^4$ is the Euclidean
Dirac operator with a chemical potential $\mu$.
In general the measure of dense QCD is complex, since there is no
matrix $P$ that satisfies for arbitrary gauge field $A$
\begin{equation}
M(A)=P^{-1}M(A)^{\dagger}P.
\end{equation}

However, we claim that the complexity of the measure is due to
fast modes, whose frequency is larger than the chemical potential,
$\omega>\mu$. If we are interested in Fermi surface phenomena or
low energy dynamics of dense matter, most of degrees of freedom in
QCD are irrelevant. For instance, modes in the deep Dirac sea are
hard to excite at low energy due to Pauli blocking by the states
in the Fermi sea and thus decoupled to physics near the Fermi
surface. On the other hand, modes near the Fermi surface are easy
to excite, since it does not cost much energy to put in or remove
the modes near the Fermi surface.

Therefore, we need to know the energy spectra of QCD near the
Fermi surface to find out the relevant modes for
the low energy dynamics of dense QCD. This is in general very
difficult since it amounts to solving QCD. However,
the problem becomes easier at extreme density because
the typical momentum transfer in the scattering of
quarks near the Fermi surface
is quite large compared to $\Lambda_{\rm QCD}$.

Due to asymptotic freedom, the QCD interaction of modes
near the Fermi surface
can be treated perturbatively and the spectrum is determined approximately
by the energy eigenvalue equation of {\em free} Dirac particles;
\begin{equation}
\left(\vec\alpha\cdot \vec p-\mu\right)\psi_{\pm}=E_{\pm}\psi_{\pm},
\end{equation}
where $\psi_{\pm}$ are the eigenstates of $\vec\alpha\cdot\vec p$ with
eigenvalues $\pm\left|\vec p\right|$.
At low
energy ($E<\mu$), the states $\psi_+$ near the Fermi surface, $|\vec
p|\sim\mu$, are easily excited, while $\psi_-$, corresponding
to the states in the Dirac sea, are completely decoupled.
Therefore, the relevant modes for the low-energy
QCD at high density ($\mu\gg\Lambda_{\rm QCD}$) are
$\psi_+$ modes and the soft gluons.

Consider a quark near the Fermi surface and
decompose the quark momentum into the Fermi momentum and a
residual momentum as
\begin{eqnarray}
p_{\mu}=\mu v_{\mu}+l_{\mu},\quad |l_{\mu}|<\mu,
\nonumber
\end{eqnarray}
where $v_{\mu}=(0,\vec v_F)$ and $\vec v_F=\vec p_F/\mu$ is the Fermi velocity,
neglecting the quark masses.
In the leading approximation in $1/\mu$ expansion,
the energy of the quark near the Fermi surface depends only on
the residual momentum parallel to the Fermi velocity, while the
perpendicular component, $\vec l_{\perp}$,
labels the degeneracy on the Fermi surface.
Therefore, the integration over the perpendicular component
should give the area of the Fermi surface,
\begin{equation}
\int {\rm d}^2l_{\perp}=4\pi p_F^2.
\end{equation}

Now, at low energy $E<\mu$, the Fermi velocity of the quark
near the Fermi surface
does not change under any scattering, since any change in the Fermi
velocity can be absorbed into the redefinition of the residual
momentum. So, it is convenient to define a Fermi-velocity dependent
field which carries the residual momentum only,
\begin{eqnarray}
\psi_+(\vec v_F,x)={1+\vec \alpha\cdot\vec v_F\over2}
e^{-i\mu\vec v_F\cdot \vec x}\psi(x).
\nonumber
\end{eqnarray}
Integrating out the irrelevant modes, $\psi_-$ and hard gluons,
we get the high density effective theory (HDET) of QCD
for dense quark matter at low energy~\cite{Hong:2000tn}.

\section{Positivity at asymptotic density}
The effective theory
is described by
\begin{equation}
\label{treeL} {L}_{\rm eff}=
\bar\psi_+i\gamma_{\parallel}^{\mu}D_{\mu}\psi_+
(\vec v_F,x)-{c_1\over2\mu}\bar\psi_+
\gamma^0({D\!\!\!\!/}_{\perp})^2\psi_+ ~+~ \cdots,
\end{equation}
where $c_1$ is a dimensionless constant due to loop effects of the
irrelevant modes and the ellipsis denotes the higher order
terms in $1/\mu$ expansion.

We note that
the Dirac operator of the effective theory in Euclidean space
is related to its hermitian conjugate by a similarity transformation,
\begin{equation}
M_{\rm eft}=~\gamma^{E}_{\parallel}\cdot D(A)~=~
\gamma_5M_{\rm eft}^{\dagger}\gamma_5.
\end{equation}
Therefore, HDET has a positive measure
in the leading order. Since the next-to-leading term is hermitian, while
the leading term is anti-hermitian, the sign problem
comes in at the next-leading order. However, the sign problem is
suppressed by $1/\mu$.

To implement HDET on lattice, it is
useful to introduce an operator formalism in which the velocity is
realized as an operator,
\begin{equation}
\label{velo} \vec{v} =   \frac{-i }{\sqrt{- \nabla^2}}
~\frac{\partial}{\partial \vec{x}}~~,
\end{equation}
since one needs to know the Fermi velocity for a given
configuration of quarks.
Then, the quasi-quarks near the Fermi surface are described by
\begin{equation}
\psi_+ = \exp \left( - i \mu x \cdot v ~ \alpha \cdot v \right)
\psi\,.
\end{equation}
Now, the effective Lagrangian density becomes
\begin{eqnarray}
\label{leading2}
{ L}_+ =  \bar{\psi}_+  \gamma^\mu_\parallel
\left(\partial^\mu + i A^\mu_+ \right) \psi_+ ~,
\end{eqnarray}
where
$A^\mu_+  =  e^{-iX} ~ A^\mu ~e^{+iX}$ and $X=\mu x \cdot v ~ \alpha \cdot v$.
Note that $\gamma^\mu_{\parallel} \partial^\mu = \gamma^\mu \partial^\mu$,
since $v \cdot \partial \, v \cdot \gamma =
\partial \cdot \gamma~$.

The partition function of dense QCD can be rewritten as
\begin{equation}
Z(\mu)=\int {\rm d}A_+~\det M_{\rm eff}(A_+)e^{-S_{\rm eff}(A_+)},
\end{equation}
and the effective action is given as
\begin{eqnarray}
S_{\rm eff}(A)=\int{\rm
d}^4x_E\left({1\over4}F_{\mu\nu}^aF_{\mu\nu}^a +{M^2\over
16\pi}\sum_{\,\vec v_F}A_{\perp\mu}^{a}A_{\perp\mu}^{a}
+\cdots\right),
\end{eqnarray}
where  $A_{\perp}=A-A_{\parallel}$, $M=\sqrt{N_f/(2\pi^2)}g_s\mu$ is
the Debye mass, and the ellipsis denotes terms
suppressed by $1/\mu$.
Therefore, we see that at the leading order HDET
has a positive measure and the lattice calculation is
possible~\cite{Hong:2002nn}.

To estimate the size of the higher-order contributions,
we calculate the correction to the vacuum energy by the naive
dimensional analysis~\cite{Manohar:1983md}. We found
\begin{eqnarray}
{\delta E_{\rm vac}\over E_{\rm vac}}\sim{\alpha_s\over 2\pi}{\Lambda\over \mu},
\end{eqnarray}
where $\Lambda\simeq\Lambda_{\rm QCD}$ is the energy scale that
we are interested in.
Therefore, the positivity of HDET is good,
as long as the chemical potential is much larger than
$\Lambda_{\rm QCD}$.

As an application of the positivity of QCD at asymptotic density,
one can establish a rigorous inequality like the Vafa-Witten
theorem~\cite{Vafa:tf} to show that the color-flavor locked (CFL)
phase~\cite{Alford:1998mk} is in fact
exact at asymptotic density.

Consider the correlator of the $SU(3)_V$ flavor currents
\begin{eqnarray}
\left<J_{\mu}^A(\vec v_F,x)J_{\nu}^B(\vec v_F,y)\right>^A
=-{\rm Tr}\,\gamma_{\mu}T^A S^A(x,y;\Delta)\gamma_{\nu}T^B
S^A(y,x;\Delta),
\end{eqnarray}
where $J_{\mu}^A(\vec v_F,x)
=\bar\psi_+(\vec v_F,x)\gamma_{\mu}T^A\psi_+(\vec v_F,x)$ and we have
introduced an infrared cut-off $\Delta$, which breaks the $U(1)$ baryon
number symmetry~\footnote{Note that any infrared regulator has to break the $U(1)$
baryon number to open a gap at the Fermi surface}.
The anomalous propagator can be rewritten as
\begin{eqnarray}
S^A(x,y;\Delta)=\left<x\right|{1\over M}\left|y\right>=\int_0^{\infty} {\rm d}\tau
\left<x\right|e^{-i\tau (-iM)}\left|y\right>
\end{eqnarray}
where $D=\partial+iA$ and
\begin{eqnarray}
M= \gamma_0\pmatrix{D\cdot V &
       \Delta \cr
\Delta & D\cdot\bar V\cr},
\end{eqnarray}
with $V=(1,\vec v_F)$, $\bar V=(1,-\vec v_F)$.
Since the eigenvalues of $M$ are bound from the below by $\Delta$, we
have the following inequality:
\begin{eqnarray}
\left|\left<x\right|{1\over M}\left|y\right>\right|
\le \int_R^{\infty}\!\!\!{\rm d}\tau \,e^{-\Delta\tau}\sqrt{\left<x|x\right>}
\sqrt{\left<y|y\right>}={e^{-\Delta R}\over\Delta}
\sqrt{\left<x|x\right>}\sqrt{\left<y|y\right>}.
\end{eqnarray}
Since the measure of HDET is positive, the vector current correlator falls off
exponentially even after integrating over the gauge fields. Therefore,
there is no Nambu-Goldstone mode along the vector channel. Combining this
with the Cooper theorem, we prove that
the CFL phase is exact.

In conclusion, we have shown that dense QCD is positive at asymptotic density.
Furthermore, a lattice calculation should be possible
using HDET, an effective theory for quasi-quarks near the Fermi surface,
as long as $\mu\gg\Lambda_{\rm QCD}$. As a consequence of the positivity,
we were able to show that the (global) vector symmetries except the
$U(1)$ baryon number
are not broken in QCD at asymptotic density.

\vskip 2in

\section*{Acknowledgments}
I would like to thank Steve Hsu for the collaboration on which this talk is
based.
This work is supported in part by the KOSEF
grant number 1999-2-111-005-5 and also  by the academic research
fund of Ministry of Education, Republic of Korea, Project No.
BSRI-99-015-DI0114.

\end{document}